\documentstyle[prl,aps,multicol]{revtex}
\input{epsf}
\begin{document}
\draft
\title{Hundred photon microwave ionization of Rydberg atoms in a static
electric field}
\author{Giuliano Benenti$^{(a,b,c)}$, Giulio Casati$^{(a,b,c)}$ and 
Dima L. Shepelyansky$^{(d,*)}$}
\address{$^{(a)}$International Centre for the Study of Dynamical Systems,
Via Lucini 3, 22100 Como, Italy}
\address{$^{(b)}$Istituto Nazionale di Fisica della Materia, 
Unit\`a di Milano, Via Celoria 16, 20133 Milano, Italy}
\address{$^{(c)}$Istituto Nazionale di Fisica Nucleare, Sezione di Milano,
Via Celoria 16, 20133 Milano, Italy} 
\address {$^{(d)}$ Laboratoire de Physique Quantique, UMR C5626 du CNRS, 
Universit\'e Paul Sabatier, F-31062 Toulouse, France}

\date{30 April 1997}
\maketitle
\begin{abstract}
We present analytical and numerical results for the 
microwave excitation of nonhydrogenic atoms in a static electric field
when up to 1000 photons are required to ionize an atom.
For small microwave fields, dynamical localization in photon number
leads to exponentially small ionization while above quantum delocalization
border ionization goes in a diffusive way. For alkali atoms 
in a static field the ionization border
is much lower than in hydrogen due to internal chaos.
\end{abstract}

\begin{multicols}{2}
\narrowtext

After two decades of investigations initiated by the pioneer experiment
of Bayfield and Koch in 1974 \cite{BK74}, the main features of the
microwave ionization of highly excited
hydrogen atoms are well understood \cite{IEEE,Jensen,Koch}.
Generally, this ionization takes place due to emergence of chaos, in
the corresponding classical problem, above a certain microwave intensity
threshold.
In the chaotic regime, quantum excitation proceeds via a diffusive absorption 
and reemission of field photons, which eventually leads to ionization.
For a real atom the quantum process can be either close to the classical 
diffusion, or strongly suppressed when the field $\epsilon$ 
is below a quantum delocalization border
$\tilde{\epsilon}_q$. This suppression is due 
to dynamical localization of
classical chaos which is produced by quantum interference. In the localized case
the distribution over the levels drops exponentially with the photon number.
This dynamical localization is analogous to
the Anderson localization in quasi one--dimensional disordered solids,
with the photon number playing the role of lattice index. The quantum
delocalization takes place if the localization length
$\ell_\phi >  N_I$, where, in atomic units,
$N_I=1/(2 n_0^2\omega)$ is the 
number of photons required for ionization, with $n_0$ being the initial 
principal quantum number and $\omega$ the field frequency.
\newline
\indent
An absolutely different scenario for the microwave ionization of alkali Rydberg
atoms had been proposed by Gallagher {\it et al.} based on experimental
results in the regime 
with small rescaled frequency $\omega_0 = \omega n_0^3 \ll 1$ 
\cite{Gall1,Gall2}.
According to this scenario, ionization appears due to 
a chain of consecutive  Landau--Zener transitions
between nearby levels, which eventually brings the electron into the continuum.
In the presence of quantum defects the Stark manifolds exhibit a structure of 
avoided crossings only for sufficiently strong fields with $\epsilon >
1/n_0^5$. 
Indeed, the  experiments \cite{Gall1,Gall2} were in agreement with the 
$\epsilon \sim 1/n_0^5$  dependence 
for the microwave ionization threshold and not with the static field border
$\epsilon_s = 1/(9 n_0^4)$. This was considered as an experimental confirmation
of the above scenario. However, this picture doesn't explain in fact how 
the propagation actually occurs via the chain of these transitions and 
why the overlapping of two nearby
levels guarantees that the electron will pass through the whole chain.
The comparison between the thresholds for hydrogen \cite{Koch} and for alkali 
atoms \cite{Gall1,Gall2,Walther} clearly shows that the border 
is lower in the latter case
and therefore it is related to the quantum defects $\delta_l$ of alkali atoms. 
Since these defects are different from zero only for orbital momentum
$l < 3$, the situation is purely quantum and  cannot be treated
by the quasi--classical approach used for hydrogen atom \cite{IEEE}.
This is the reason why so long after experiments have been made 
\cite{Gall1,Gall2} no detailed theory has been developed.
To demonstrate a theoretical difficulty we note that hundreds of photons are
required to ionize atoms in this regime ($N_I=n_0/2\omega_0=300$ for $n_0=60$
and $\omega_0=0.1$).
\newline
\indent
In this Letter we propose another mechanism for the microwave ionization of 
Rydberg atoms in a static electric field
which is qualitatively different from the Gallagher {\it et al.} scenario.
We argue that ionization in this case
is not due to Landau--Zener transitions but
to a quantum diffusive excitation in energy $E$ (or in the photon
number $N=E/\omega$). Such quantum diffusion in the low frequency regime 
$1/n_0 < \omega_0 \ll 1$ becomes possible due to appearence of quantum chaos
for Rydberg atoms in a static electric field. Indeed, a
recent theoretical study \cite{Kleppner} 
showed that the level spacing
statistics $P(s)$ in such atoms, for a sufficiently strong static field, 
is described by the Random Matrix Theory. 
These results indicate also a chaotic structure of eigenstates. 
This internal quantum chaos can lead to a diffusive excitation in energy 
even for
a quite weak microwave field. In this respect, the situation is different from
the hydrogen atom where chaos could appear only above some classical field
threshold. 
Another consequence
of internal chaos is an increase of density of effectively coupled
states $\rho_c$. This gives a larger localization length $\ell_q$ 
($\ell_q \propto
\rho_c$) and therefore significantly decreases the delocalization border 
as compared to the
hydrogenic case. A similar effect of chaotic enhancement of localization length
due to internal chaos has been studied recently for hydrogen atoms in 
magnetic and microwave fields \cite{mag1,mag2}.

\begin{figure}
\vglue 0.2cm
\epsfxsize=3.3in
\epsfysize=2.5in
\epsffile{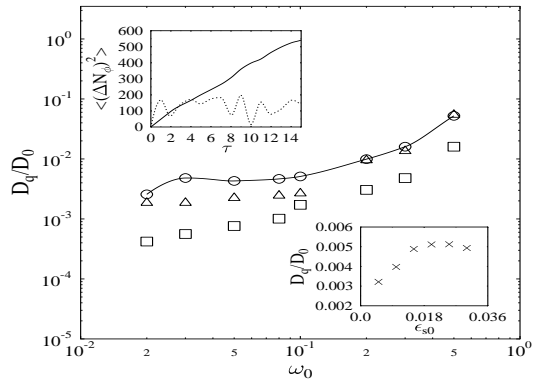}
\vglue 0.cm
\caption{Diffusion rate ratio $D_q/D_0$ as a function of microwave
frequency $\omega_0$
for Rb (circles), Na (triangles) and Li (squares) at $n_0=60$, 
$\epsilon_{s0}=0.02$. The full line is drawn to guide the eye. 
The upper insert gives 
$<(\Delta N_\phi)^2>$ vs. $\tau$ at $\epsilon_0=0.03$, $\omega_0=0.1$: 
diffusive excitation for Rb (full curve $D_q/D_0=0.005$) 
and small oscillations for hydrogen
(dotted curve with 100 times magnification).
The lower insert illustrates the weak dependence of $D_q/D_0$ on $\epsilon_{s0}$
for Rb at $\omega_0=0.1$, $\epsilon_0 \geq 0.015$ (crosses).} 
\label{fig1}
\end{figure}

To check the above picture of quantum photonic diffusion, we numerically studied
the excitation of alkali Rydberg atoms (Rb, Na, Li)
in a static electric field $\epsilon_s$ and parallel, linearly polarized,
microwave field $\epsilon \sin \omega t$, for magnetic quantum number $m=0$.
We chose the rescaled value $\epsilon_{s0}=\epsilon_s n_0^4
\approx 0.02$ so that the statistics $P(s)$ 
for levels with $55 \leq n_0 \leq 72$ in Rb and Na
was close to the RMT results. At the same time $P(s)$ for
Li was closer to Poisson statistics due to a smaller 
value of quantum defects
\cite{mag3}. The investigation of time evolution, in the eigenbasis of the
unperturbed problem ($\epsilon=0$), showed that an initial eigenstate with 
energy $E_0$ spreads diffusively over the unperturbed energies $E_\lambda$, 
namely the square variance of the photon number
$\sigma=<(\Delta N_\phi)^2>=<(E_\lambda-E_0)^2>/\omega^2$ 
initially grows linearly with 
the number of microwave periods $\tau$.  
Our quantum simulation allows to determine the value of quantum diffusion 
rate in energy per unit time $D_q=<(\Delta E)^2>/\Delta t$. This rate
$D_q$ can be compared with the diffusion rate in 
hydrogen at $\omega_0=1$, given by 
$D_0=\epsilon^2 n_0/2$ \cite{IEEE}.
The dependence of the ratio $D_q/D_0$ on the frequency
$\omega_0$, for rescaled fields $0.005 \le \epsilon_0 \le 0.03$
($\epsilon_0=\epsilon n_0^4$), is shown in Fig. 1.

In our computations we consider the most interesting
frequency region $\omega_0 \leq 0.5$ where ionization 
in hydrogen is close to the static field ionization.
In our opinion the ratio $D_q/D_0$ in Fig. 1 is small 
because collisions with the core, which are responsible for 
diffusion, happen rarely, with a frequency
of classical precession in $l$, which is $\omega_{s0} =3 \epsilon_{s0} \ll 1$. 
Instead, for $\omega_0 >1$
the ratio $D_q/D_0 \approx 1/3 \omega_0^{4/3}$ 
is similar to the hydrogen case
\cite{mag3}. 
Notice that the dependence of $D_q/D_0$ on $\omega_0$ is 
very flat in the interval
$0.02 < \omega_0 < 0.3$. The origin of this fact is not quite clear
since asymptotically ,for small $\omega_0$, one should expect 
$D_q \propto \omega_0^2$ 
\cite{mag1,mag2}. 
(Apparently, this plateau in frequency appears due to
one-photon transitions between levels in the Stark multiplet. 
As a result its rescaled size is expected to be
$\omega_{s0} \approx 3 \epsilon_{s0}$, which is in approximate agreement 
with the data of Fig.1.)
We also checked that a change of $n_0$ from $60$ to $28$
at the same classically scaled parameters
did not affect the ratio $D_q/D_0.$

\begin{figure}
\vglue 0.2cm
\epsfxsize=3.3in
\epsfysize=2.5in
\epsffile{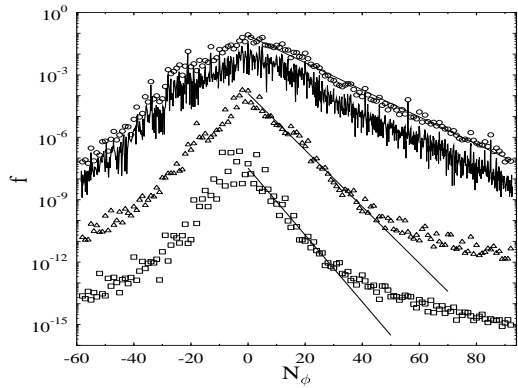}
\vglue 0.cm
\caption{Probability distribution 
over the eigenstates at $\epsilon =0$ (full
line) and in one-photon intervals (circles) for
Rb: $n_0=60, \epsilon_{s0}=0.02, \omega_0=0.1, \epsilon_0 = 0.005,$
$180 \leq \tau \leq 200, D_q/D_0=0.0051, \ell_q = 9.1$. The straight line shows 
the fit for exponential localization with $\ell_{qN} = 13.3$.
The one--photon probabilities are also shown, in the same conditions,
for Na (triangles,
shifted down by $10^3$, $\ell_{q}=4.8, \ell_{qN}=6.4$) and Li (squares.
shifted down by $10^6$, $\ell_{q}=3.1, \ell_{qN}=5.4$).} 
\label{fig2}
\end{figure}

The above diffusive excitation in energy induced by internal chaos 
may eventually be localized at long times due to quantum interference
effects in a way similar to photonic localization in a 
complex molecular spectrum \cite{Dima87,IEEE}. The localization
length $\ell_q$, expressed in the number of photons, is proportional to the
one--photon transition rate $\Gamma$ and to the density of coupled states
$\rho_c$: $\ell_q \sim \Gamma \rho_c$ \cite{Dima87}. 
Since $\Gamma \sim D_q /\omega^2$
we obtain 
\begin{eqnarray}
\label{lq}
\ell_q = \ell_\phi \frac {D_q}{D_0\omega_0^2} n_0.
\end{eqnarray}
Here the length $\ell_q$ is expressed via the localization length
in hydrogen $\ell_\phi=3.3\epsilon_0^2 n_0^2$ at $\omega_0=1$. 
Notice that in the non-hydrogenic case
the density of states is $\rho_c=n_0^4$ due to internal chaos while in
hydrogen the effective
density is smaller, that is $\rho=n_0^3$ due to existence of an additional
integral of motion \cite{IEEE}.  The above expression for $\ell_q$
is valid for $\ell_q>1$ and $\omega \rho_c=\omega_0 n_0>1$.
The localization leads to an exponential decay of probability distribution 
$|\psi_N|^2 \sim \exp(-2 |N_\phi|/\ell_q)$ in the photon number 
$N_\phi=(E_\lambda-E_0)/\omega$.

\begin{figure} 
\vglue 0.2cm
\epsfxsize=3.3in
\epsfysize=2.5in
\epsffile{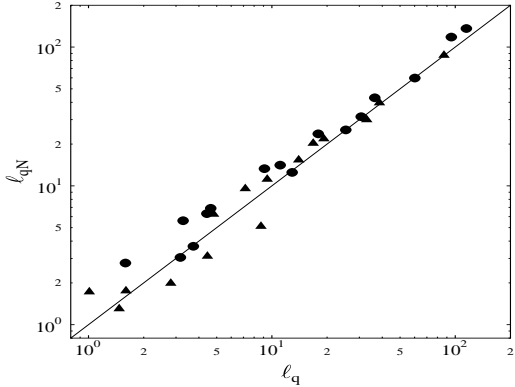}
\vglue 0.cm
\caption{The numerically computed localization length $\ell_{qN}$
vs. theoretical $\ell_q$ (Eq. (1))  for Rb (points) and Na (triangles)
at $n_0=60, \epsilon_{s0}=0.02$ and $0.02 \leq \omega_0 \leq 0.5$,
$0.003 \leq \epsilon_0 \leq 0.01$. The straight line gives 
$\ell_{qN}=\ell_q$, while the numerical average is 
$<\ell_{qN}/\ell_q>=1.17 \pm 0.28$.
} 
\label{fig3}
\end{figure}

In order to check the theoretical prediction (1) we numerically 
computed the quantum evolution following it up to 200 microwave periods.
The probability distribution $f_\lambda$ over the eigenstates
of the static field problem ($\epsilon=0$) with energies $E_\lambda$, 
was averaged over 10 - 20 periods to suppress the fluctuations.
An initial state at $\epsilon_{s0}=0.02$
was chosen as an eigenstate with
energy $E_{\lambda_0} \approx E_0 = 1/2n_0^2$ and $n_0=60$.
The system parameters were varied in the intervals:
$0.02 \leq \omega_0 \leq 0.5$, $0.003 \leq \epsilon_0 \leq 0.03$
and $60 \leq N_I \leq 1500$. The total basis included up to 1150 states.
\newline
\indent
A typical example of stationary distribution $f$ is shown in Fig. 2.
It clearly demonstrates exponential localization of diffusive excitation.
One can note that among the three cases shown (Rb, Na, Li) the most localized 
is the case of Li, for which the quantum defect is minimal 
and therefore the internal chaos is the most weak. 
Notice also that in the case of Na and Li,
for $f_N < 10^{-7}$, the probability $f_N$ starts to decay in a much slower 
way with $l_q \approx 25$ (Fig. 2). We attribute this effect to 
a significant modification of hydrogenic basis on highly excited
levels where a static field becomes quite strong and tunneling
effects for probability decay should be taken into account.
\newline
\indent
In order to find the localization length $\ell_q$ we first compute the total
probabilities $f_N = |\psi_N|^2$ in one-photon intervals 
$[N_\phi-1/2,N_\phi+1/2]$ around 
integer values of $N_\phi=E_\lambda/\omega$ and then extract the numerical
$\ell_{qN}$ value from the least square fit for $\ln f_N$. 
The analysis of the 
numerical data $\ell_{qN}$ shown in Fig.3 confirms the theoretical prediction
(1) for $\ell_q$ with $D_q$ rates taken from Fig.1.

Equation (1) allows to determine the quantum delocalization border
above which localization effects become unimportant and ionization 
goes in a diffusive way. This happens for $\ell_q > N_I$, which gives
the quantum delocalization border for the rescaled field $\epsilon_q=
\tilde{\epsilon}_q n_0^4$:
\begin{eqnarray}
\label{deloc}
\epsilon_q = 0.4 \omega^{1/3} \omega_0^{1/6} \sqrt{D_0/D_q}.
\end{eqnarray}
Our numerical data indeed show that above this border 
complete delocalization takes place,
contrarily to the case of hydrogen atom at the same field parameters (Fig.4).
The strong fluctuactions 
in the hydrogenic distribution
$f_N$ indicate that it is quite inhomogeneous inside the atomic shells.

\begin{figure}
\vglue 0.2cm
\epsfxsize=3.3in
\epsfysize=2.5in
\epsffile{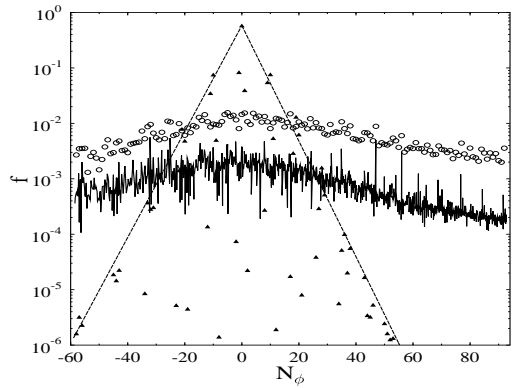}
\vglue 0.cm
\caption{Same as Fig.2 but in a delocalized case for Rb, with $\epsilon_0=0.03
> \epsilon_q=0.028$ and $40 \leq \tau \leq 50$. Full triangles give 
one--photon interval probability distribution $f_N$ for hydrogen atom
at the same system parameters (dashed lines show the distribution envelope). 
} 
\label{fig4}
\end{figure}

For fixed $\omega_0$ the delocalization border (2) 
scales as $\tilde{\epsilon_q} 
\sim 1/n_0^5$, namely it is $\sqrt{n_0}$ times smaller than for hydrogen
at $\omega_0 \sim 1$. This
drop of $\epsilon_q$ is related to the appearence of internal chaos in
nonhydrogenic atoms in a static electric field
which effectively enhances the interaction with the microwave radiation. 
According to (2) the ratio 
of $\epsilon_q$ to the ionization border proposed by Gallagher 
{\it et al.} $\epsilon_G \approx 1/n_0$ depends only on $\omega_0$:
$\epsilon_q/\epsilon_G = 0.4 \sqrt{\omega_0 D_0/D_q}$. Our data from Fig.1
indicate that this ratio varies rather weakly with $\omega_0$ in the interval
$0.02 < \omega_0 <0.5$ (see Fig.5).  In spite of the fact that in this range
$\epsilon_q \sim \epsilon_G$, 
the physical interpretation is rather different from Gallagher et al. scenario.
Indeed, the situation for $\epsilon_0 < \epsilon_q \sim 1/n_0$
is strongly nonperturbative since for 
$\epsilon_q (2\omega_0/n_0)^{1/2} < \epsilon_0$ many photons are
absorbed with $l_q > 1$.
\newline
\indent
The theory developed above allows to understand ionization of atoms 
in a static electric field. It is possible to expect that for $\omega_0 \ll 1$
the situation will remain similar even without static field
since its role will be played by a slowly varying microwave field.
However, one should be careful in extending the theory to a zero static field 
case. Indeed, our numerical results show that there, 
the probability distribution
in orbital momentum is qualitatively different.
For example, at  $\epsilon_0 \sim 0.02,
\omega_0 \sim 0.2$ only few $l$-states  are mixed while
with additional $\epsilon_{s0} \sim 0.02$ probability spreads over all
accessible $l$. The localization in $l$ space at $\epsilon_{s0}=0$
had been also discussed for Rb atoms in \cite{Blumel}. 
The physical reason of this difference is related to the fact that
the condition $\omega_0 \ll 1$ is not sufficient to treat the
microwave field $\epsilon_0$ as quasi-static. For that one should
require $\omega_0 \ll \omega_{s0} \approx 3\epsilon_0$ since 
the precession frequency $\omega_{s0}$ determines
oscillations in $l$ \cite{mag3}. However, this condition is not compatible
with the requirement $\omega_0 > 1/n_0$ for the considered region of $n_0$.
For $\epsilon_{s0}=0$ chaos is induced by the microwave field,
the distribution in $l$ is nonhomogeneous
and a detailed theory for this case becomes more complicated
as compared to the nonzero static field where internal
chaos is already present. 

\begin{figure}
\vglue 0.2cm
\epsfxsize=3.3in
\epsfysize=2.5in
\epsffile{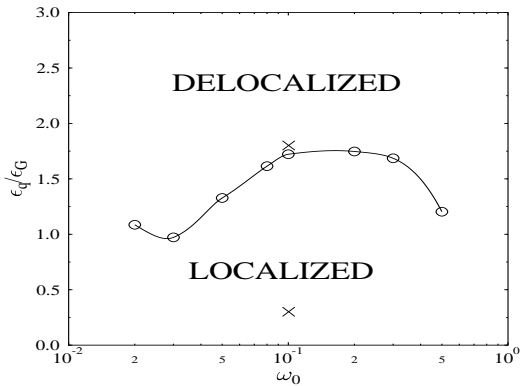}
\vglue 0.cm
\caption{The ratio of the quantum delocalization border $\epsilon_q$ to
the border $\epsilon_G=1/n_0$ vs. $\omega_0$ for
Rb (circles) obtained from Eq. (2) and data of Fig.1. The full line is
drawn to guide the eye, crosses refer to the cases of Figs. 2,4.
} 
\label{fig5}
\end{figure}

Another difficulty for direct comparison with the
experiments \cite{Gall1,Gall2} is that the latter
were mainly done in the regime $\omega_0 < 1/n_0$. 
There is only one case for Na at $n_0=28$, $\omega_0 n_0 = 0.76$
and $\epsilon_{s0} \approx 0.024$  which
is not far from our regime (Fig. 2d in \cite{Gall2}). 
Here the experiment gives
the ionization border $\epsilon_{0ex} \approx 0.002$.
This value is about $20$ times smaller than the  quantum delocalization 
border given by Eq. (2) with $D_q/D_0 = 0.0027$. 
We relate this difference to a tunneling ionization in a 
static field which becomes important for very long
interaction times ($3 \times 10^5$ Kepler periods in \cite{Gall2}).
Indeed, our preliminary numerical data  \cite{mag3}
show that there is a change of slope in the probability
decay similar to one in Fig.2 
($l_{qN} \approx 4$ for $f_N > 10^{-5}$ and 
$l_{qN} \approx 25$ for $f_N < 10^{-5}$).
This means that in this case tunneling ionization
plays a dominant role while
the microwave only slightly increases its rate. 
An increase of the principal quantum number up to $n_0=60$
strongly suppresses the tunneling in a static field
and the diffusive microwave excitation becomes dominant (Figs. 2,4).
The effects of tunneling for  $n_0 \approx 30$
will become less important for experiments with a shorter 
interaction time ($\tau \sim 100$) on which
the dynamical localization dominates.
It is also possible that for such long times as in \cite{Gall2}
some noise in a microwave signal can strongly affect ionization.
The above discussion shows that
more detailed experimental investigations of microwave ionization
in a static electric field are highly desirable.
They will allow to make a detailed test of 
dynamical localization theory in the regime when up to 1000 photons are
required to ionize an atom.

\end{multicols}

\end{document}